\documentclass[pra,twocolumn,showpacs]{revtex4}
\usepackage{amsmath}
\usepackage{mathrsfs}
\usepackage{graphicx}
\usepackage{hyperref}
\usepackage{dcolumn}
\usepackage{bm}

\begin{document}

\title{Bulk vortices and half-vortex surface modes in parity-time-symmetric media}

\author{Huagang Li$^{1,2}$}
\author{Xing Zhu$^1$}
\author{Zhiwei Shi$^3$}
\author{Boris A. Malomed$^4$}
\author{Tianshu Lai$^{1,}$}
\altaffiliation{Corresponding author. Email: stslts@mail.sysu.edu.cn}
\author{Chaohong Lee$^{1,}$}
\altaffiliation{Corresponding author. Email: chleecn@gmail.com}

\affiliation{$^1$State Key Laboratory of Optoelectronic Materials and Technologies, School of Physics and Engineering, Sun Yat-Sen University, Guangzhou 510275, China}

\affiliation{$^2$Department of Physics, Guangdong University of Education, Guangzhou
510303, China}

\affiliation{$^3$School of Information Engineering, Guangdong University of Technology,
Guangzhou 510006, China}

\affiliation{$^4$Department of Physical Electronics, School of Electrical Engineering, Faculty of Engineering, Tel Aviv University, Tel Aviv 69978, Israel}

\date{\today }

\begin{abstract}
We demonstrate that in-bulk vortex localized modes, and their surface half-vortex (\textquotedblleft horseshoe") counterparts self-trap in two-dimensional (2D) nonlinear optical systems with $\mathcal{PT}$-symmetric photonic lattices (PLs). The respective stability regions are identified in the underlying parameter space. The in-bulk states are related to truncated nonlinear Bloch waves in gaps of the PL-induced spectrum. The basic vortex and horseshoe modes are built, severally, of four and three beams with appropriate phase shifts between them. Their stable complex counterparts, built of up to 12 beams, are reported too.
\end{abstract}

\maketitle


Nonlinear spatially periodic systems support diverse types of self-trapped in-gap states. In particular, spatial gap solitons~\cite{101,102} originate from the interplay between the periodicity and nonlinearity. Further, surface gap solitons~\cite{Kartashov,Rosberg,Suntsov2006,Szameit,Wang} appear at the interface between a uniform medium and a photonic lattice (PL) built into a nonlinear material. Extended self-trapped waves~with steep edges also exist in these settings, being related to truncated nonlinear Bloch waves \cite{Anker,Alexander}. Modes of the latter type provide a link between extended nonlinear Bloch waves~\cite{Bronski,Louis} and tightly localized gap solitons~\cite{101,102,BrazhKon}.

Recently, a great deal of interest has been drawn to the realization of parity-time ($\mathcal{PT}$) symmetry in optics. Originally, this concept was developed in quantum mechanics, where it was demonstrated that, beyond the conventional Hermitian Hamiltonians, their $\mathcal{PT}$-symmetric non-Hermitian counterparts may also give rise to purely real (hence physically relevant) spectra~\cite{Bender1,Bender2,Bender3,Bender4}. Following the similarity between quantum mechanics and paraxial optics~\cite{Longhi2009,Szameit2010}, $\mathcal{PT}$-symmetric optical systems with complex refractive indices~\cite{Kottos2010,Yin2013} have been extensively studied theoretically~\cite{Klaiman,El-Ganainy,Sukhorukov,Radik,Barash, Ramezani, Abdullaev,Makris,Longhi1,Musslimani,Hang2013,Lee2013, Lumer2013} and experimentally~\cite{Guo,Ruter,Bittner,Regensburger,Feng2012, Regensburger2013,Wimmer2013,Peng2013}. In this context, $\mathcal{PT}$-symmetric PLs play an important role. Taking into account the non-orthogonality of the respective eigenmodes, their coupled-mode description had to be reformulated via the variational principle~\cite{El-Ganainy}. The light propagation in $\mathcal{PT}$-symmetric PLs embedded into linear media were analyzed preliminarily~\cite{Makris,10-0}. Further, it has been found that 1D and 2D spatial gap solitons exist in PLs built into a nonlinear material~~\cite{Musslimani,1301,1302,1303,1304,18-0}.

Although Bloch waves~\cite{El-Ganainy} and gap solitons~\cite{Musslimani,1301,1302,1303,1304,18-0} were studied before in the context of some $\mathcal{PT}$-symmetric PLs, the comprehensive study of self-trapped states in 2D $\mathcal{PT}$-symmetric systems combining lattices and nonlinearity was not reported yet. In particular, such self-trapped nonlinear states may serve as a necessary link between spatially localized gap solitons and extended nonlinear Bloch waves under the $\mathcal{PT}$ symmetry.

Self-trapped vortices and surface modes are of great interest in the context of the $\mathcal{PT}$-symmetric settings. Indeed, the study of nonlinear surface modes pinned on the interface of a $\mathcal{PT}$-symmetric system opens a way to explore the interplay between surface effects, the nonlinearity, and the $\mathcal{PT}$-symmetry. On the other hand, the analysis of localized vortices supported by $\mathcal{PT}$-symmetric PLs should shed light on the cooperation and competition of the $\mathcal{PT}$-symmetry with the azimuthal instability and spatial periodicity.

In this work, we show the existence of in-bulk and surface self-trapped states in 2D nonlinear systems with $\mathcal{PT}$-symmetric PLs. In particular, we report in-bulk solitary vortices and novel half-vortex surface modes. Stable half-vortex surface modes appear as \textquotedblleft horseshoes" pinned on the interface between a uniform linear medium and a nonlinear medium with built-in $\mathcal{PT}$-symmetric PL. The in-bulk vortices and surface \textquotedblleft horseshoes" have a common linear stability region at intermediate values of propagation constants.

We consider the light propagations in two nonlinear systems with $\mathcal{PT}$-symmetric PLs: a uniform setting of the nonlinear $\mathcal{PT}$-symmetric PL, and a composite setting of the nonlinear $\mathcal{PT}$-symmetric PL at the left side ($x<0$) and a uniform linear medium at the right side ($x>0$). Assuming that the light propagates along the $z$-axis, the amplitude of the electromagnetic field is written as $E(x,y,z,t)=E(x,y,z)e^{i(\Gamma z-\omega t)}$, with carrier wavenumber $\Gamma $ and frequency $\omega $. With the effective refractive index including contributions from the complex PL and the Kerr effect, $n^{\mathrm{PL}}=n_{0}^{\mathrm{PL}} +n^{\mathrm{R}}(x,y)+in^{\mathrm{I}}(x,y) +n^{\mathrm{NL}}|E|^{2}$, the amplitude obeys the nonlinear Schr\"{o}dinger equation with the complex potential,
\begin{eqnarray}
&&2i\Gamma \frac{dE}{dz}+\frac{\partial ^{2}E}{\partial x^{2}} +\frac{\partial ^{2}E}{\partial y^{2}}+\left[ \left( k_{0}n_{0}^{\mathrm{PL}}\right) ^{2}-\Gamma ^{2}\right] E  \notag \\
&&+2k_{0}^{2}\left[ n_{0}^{\mathrm{PL}}\left( n^{\mathrm{R}}+in^{\mathrm{I}}\right) +n_{0}^{\mathrm{PL}}n^{\mathrm{NL}}|E|^{2}\right] E=0.
\label{eq:one}
\end{eqnarray}%
Here, $k_{0}=\omega /c$ is a constant, $n_{0}^{\mathrm{PL}}$ is the background refractive index, $n^{\mathrm{R}}(x,y)$ and $n^{\mathrm{I}}(x,y)$ are real and imaginary (gain/loss) parts of the spatial modulation of the local index, and $n^{\mathrm{NL}}$ is the Kerr coefficient. Similarly, the light propagation in the linear uniform medium obeys the paraxial equation
\begin{equation}
2i\Gamma \frac{dE}{dz}+\frac{\partial ^{2}E}{\partial x^{2}}+\frac{\partial ^{2}E}{\partial y^{2}}+\left[ \left( k_{0}n_{0}^{\mathrm{lin}}\right)^{2}-\Gamma ^{2}\right] E=0,  \label{eq:two}
\end{equation}%
with the respective real refractive index, $n_{0}^{\mathrm{lin}}$.

\begin{figure}[tbh]
\centerline{\includegraphics[width=1.0\columnwidth]{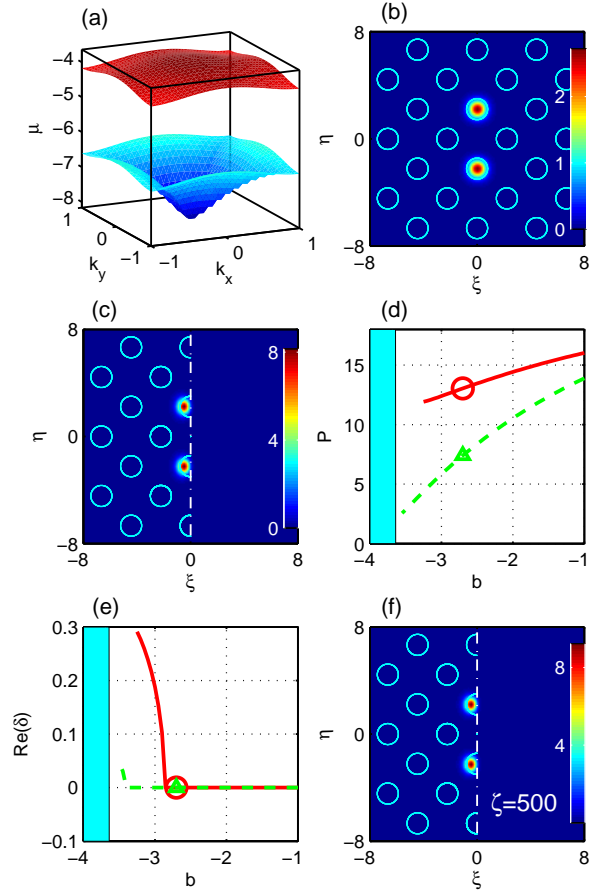}}
\caption{(Color online) (a) The bandgap structure for the 2D $\mathcal{PT}$-symmetric photonic lattice: $\protect\mu $ is the propagation constant, and $k_{x}$, $k_{y}$ are Bloch wavenumbers in $\protect\xi$ and $\protect\eta $ directions. (b,c) Intensity profiles of self-trapped modes for propagation constant $b=-2.70$ in the uniform and truncated systems, respectively. The white dot-dash line depicts the interface in the truncated system. (d) Power $P$ and (e) the real part of the instability growth rate, $\mathrm{Re}(\protect\delta )$, of the self-trapped modes versus the propagation constant, $b$. Green triangles and red circles in (d) correspond to the modes shown in (b) and (c), respectively. The green dashed and red solid lines in (e) represent, severally, in-bulk and surface self-trapped modes. (f) The density profile at $\protect\zeta =500$, evolved from the initial self-trapped mode (c) with $5\%$ noise. Parameters are $\protect\theta =0.1$, $V_{0}=-5$ and $G=-20$.}
\label{fig:one}
\end{figure}

We normalize the equations by defining $\zeta =\left( 2\Gamma
w_{0}^{2}\right) ^{-1}z$, $\xi =x/w_{0}$, $\eta =y/w_{0}$, and $q=\sqrt{n_{0}^{\mathrm{PL}}/\left(2n^{\mathrm{NL}}\right)} \left(w_{0}k_{0}n_{0}^{\mathrm{PL}}\right)^{-1} E e^{\frac{i}{2\Gamma} \left[(k_{0}n_{0}^{\mathrm{PL}})^{2}-\Gamma ^{2}\right] z}$ with an arbitrary scaling factor $w_{0}$. The accordingly rescaled form of Eqs.~(\ref{eq:one}) and (\ref{eq:two}) is%
\begin{gather}
i\frac{dq}{d\zeta }+\nabla _{\perp }^{2}q+R(\xi ,\eta)q +|q|^{2}q=0, \label{eq:seven} \\
i\frac{dq}{d\zeta }+\nabla _{\perp }^{2}q+Gq=0,  \label{eq:eight}
\end{gather}%
with $\nabla _{\perp }^{2}=\partial ^{2}/\partial \xi^{2} +\partial^{2}/\partial \eta^{2}, R=2n_{0}^{\mathrm{PL}}k_{0}^{2}w_{0}^{2}\left( n^{\mathrm{R}}+in^{\mathrm{I}}\right)$, and $G=w_{0}^{2}k_{0}^{2}\left[ \left(n_{0}^{\mathrm{lin}}\right)^{2} -\left(n_{0}^{\mathrm{PL}}\right)^{2}\right]$. The complex $\mathcal{PT}$-symmetric potential, $R(\xi ,\eta)\equiv V(\xi ,\eta )+iW(\xi ,\eta )$, is chosen as%
\begin{eqnarray*}
V(\xi ,\eta ) &=&V_{0}\left[ \cos^2 \left( \frac{\eta -\xi }{\sqrt{2}} \right) +\cos^2 \left(\frac{\eta +\xi}{\sqrt{2}} \right) \right] , \\
W(\xi ,\eta ) &=&\theta \{\sin [\sqrt{2}(\eta -\xi )]+\sin [\sqrt{2}(\eta+\xi )]\},
\end{eqnarray*}%
with amplitudes $V_{0}$ and $\theta $ of the modulation of the real and imaginary parts of the refractive index. This PL is a $45^{\circ}$ counterclockwise rotation of the one considered in Refs.~\cite{Musslimani,Makris}. The configurations of PLs are shown by the white-blue circles in the $(\xi, \eta)$-plane. Its band-gap structure can be derived by using the plane-wave expansion method based on the Floquet-Bloch theorem, see Fig.~\ref{fig:one}~(a).

To combine Eqs.~(\ref{eq:seven}) and (\ref{eq:eight}) into a single equation, we define a step function, $U(\xi )=1$ at $\xi <0$ and $U(\xi )=0$ at $\xi >0$:
\begin{eqnarray}
i\frac{dq}{d\zeta }&+&\nabla _{\perp }^{2}q+U(\xi )R(\xi ,\eta)q \nonumber\\
&+&\left[1-U(\xi)\right] Gq+U(\xi )|q|^{2}q=0.  \label{eq:jia_1}
\end{eqnarray}%
The stationary solution with real propagation constant $b$ is looked for as $q(\xi ,\eta ,\zeta )=u(\xi ,\eta)e^{ib\zeta}$, where complex function $u(\xi ,\eta )$ obeys equation
\begin{eqnarray}
\nabla _{\perp }^{2}u+U(\xi )R(\xi ,\eta )u &+&\left[1-U(\xi)\right] Gu \nonumber\\
&+&U(\xi)|u|^{2}u-bu=0.\label{eq:five}
\end{eqnarray}

To find the stationary self-trapping solutions, we used numerical simulations with the modified squared-operator method~\cite{17-0}. While the existence and stability of the simplest single-beam solitons in the present setting is quite evident, as the first step of the analysis we produce double-beam self-trapped states. For $b=-2.70$ and $G=-20$, the in-bulk and surface double modes are displayed in Fig.~\ref{fig:one}(b,c). Due to the presence of the interface, the intensity of the surface self-trapped states is larger than the in-bulk ones, at the same propagation constant. The dependence of the total power, $P=\int \int |u \left(\xi,\eta\right) |^{2}d\xi d\eta$, on the propagation constant $b$ demonstrates that the power of the surface modes is also larger than that of the in-bulk ones, see Fig.~\ref{fig:one}(d). Different from the single-beam solitons~\cite{18-0}, both surface and in-bulk self-trapped states in the semi-infinite gap  do not exist near the first Bloch band in Fig.~\ref{fig:one}(d).

Stability of the self-trapped modes was investigated by means of the linearization for small perturbations. To a given stationary state, $q_{0}(\xi ,\eta )=u(\xi ,\eta )e^{ib\zeta }$, the perturbation is added as $q_{1}(\xi ,\eta )=\epsilon \lbrack F(\xi ,\eta )e^{\delta \zeta }+G^{\ast}(\xi ,\eta)e^{\delta ^{\ast }\zeta }]e^{ib\zeta }$ with infinitesimal $\epsilon $~\cite{Musslimani,18-0}, where $F(\xi ,\eta )$ and $G(\xi,\eta)$ are two perturbation eigenfunctions, $\delta$ is the corresponding growth rate, and the star ($\ast$) stands for the complex conjugate. From Eq.~(\ref{eq:jia_1}), the following linearized equations are derived:
\begin{eqnarray}
-i\delta F&=&\left[\nabla_{\perp}^{2}+U(\xi)R(\xi,\eta)+\psi +2U(\xi)|u|^{2}\right]F \nonumber\\ &&+U(\xi )u^{2}G, \\
+i\delta G&=&\left[ \nabla _{\perp }^{2}+U(\xi )R^{\ast} (\xi,\eta)u+\psi+2U(\xi )|u|^{2}\right]G \nonumber\\ &&+U(\xi)\left(u^{\ast}\right)^{2}F,
\end{eqnarray}
with $\psi(\xi,\eta)\equiv \left[1-U(\xi)\right]G-b$. As usual, the self-trapped mode is linearly unstable if there is an eigenvalue with $\mathrm{Re}(\delta )>0$. As seen in Fig. \ref{fig:one}(e), the two-beam in-bulk and surface self-trapped modes have a common stable region, with $\mathrm{Re}(\delta)=0$, at intermediate values of propagation constants $b$. The predicted stability of the modes has been verified in direct simulations of Eq. (\ref{eq:jia_1}) with $5\%$ random noise added as an initial perturbation, see an example for $b=-2.70$ in Fig.~\ref{fig:one}(f).

\begin{figure}[tbh]
\centerline{\includegraphics[width=1.0\columnwidth]{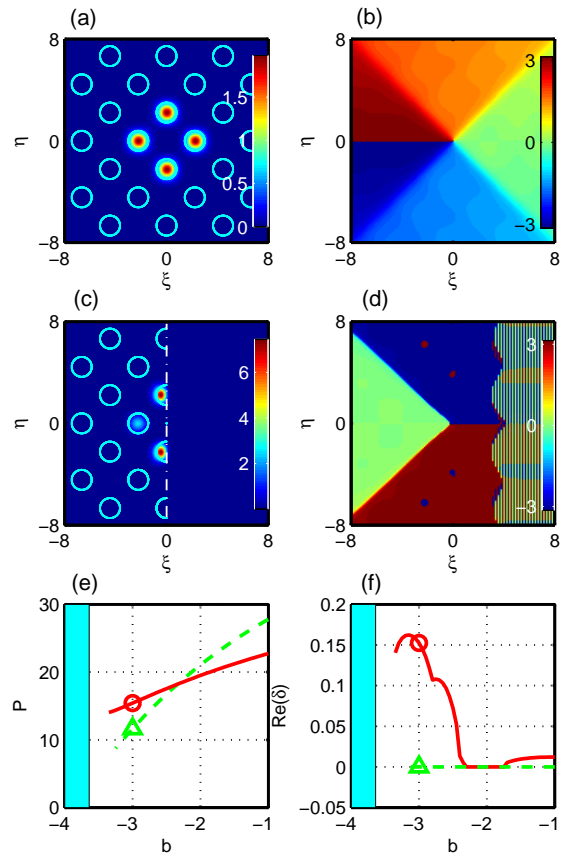}}
\caption{(Color online) (a,b) The intensity profile of the in-bulk solitary vortex, and the corresponding phase distribution, for propagation constant $b=-3.00$. (c,d) The intensity profile and phase distribution for the three-beam surface self-trapped state at $b=-3.00$. (e) Power $P$ of the self-trapped states versus $b$. (f) The real part of the instability growth rate, $\text{Re}(\protect\delta )$, versus $b$. Green triangles and the red circles correspond to the modes shown in (a) and (c), respectively. Parameters are $\protect\theta =0.1$, $V_{0}=-5$, and $G=-20$. }
\label{fig:two}
\end{figure}

Adding more beams with phase shifts between them, one can construct composite vortices. For an example, a composite vortex with the total phase circulation of $2\pi$ may appear as a four-beam complex with the off-site vortex core in the center and  the phase shifts $\pi/2$ between adjacent beams~\cite{YangOL,BBB}. We have found that the composite vortices can exist in the system of the uniform setting. The intensity profile and phase distribution of a typical stable four-beam vortex in the uniform setting system are shown in Fig.~\ref{fig:two}(a) and (b) for $b=-3.00$. Near the interface in of the composite setting system, there are no complete vortex modes, while there appear essentially new surface modes, in the form of \emph{half-vortices} (\textquotedblleft horseshoes"), built of three beams, see Figs.~\ref{fig:two}(c) and (d). The dependence of the power $P$ on the propagation constant $b$ shows that, although the half-vortex mode (c) is built of three beams, its power $P$ is larger (near the first Bloch band) than that of the in-bulk vortex mode (a), which is composed of four beams, see Fig.~\ref{fig:two}(e). The linear stability analysis shows that there exists a common stability region at intermediate values of propagation constants $b$ for the in-bulk vortices and surface \textquotedblleft horseshoes", see Fig.~\ref{fig:two}(f).

\begin{figure}[tbh]
\centerline{\includegraphics[width=1.0\columnwidth]{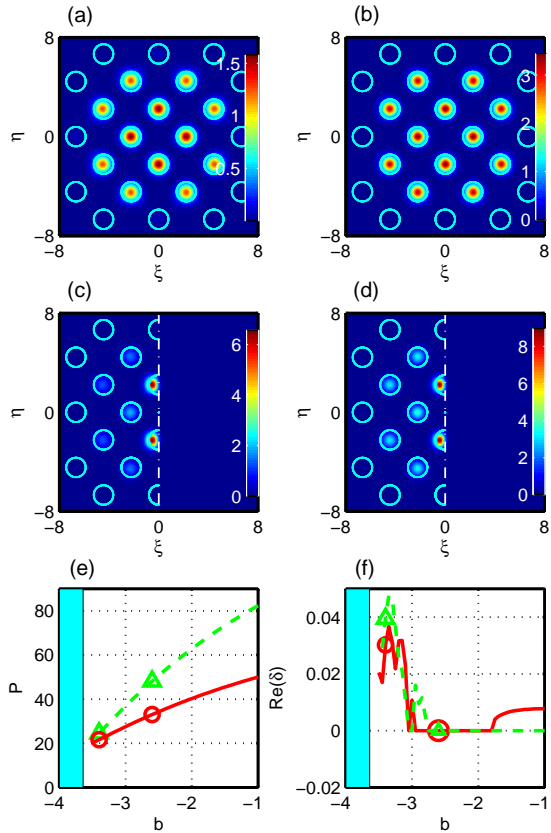}}
\caption{(Color online) Intensity profiles of the 12-beam in-bulk self-trapped states at different propagation constants: (a) $b=-2.60$, $b=-3.40$ (b). Profiles of the 7-beam surface states: (c) at $b=-2.60$, (d) at $b=-3.40$. (e) Power $P$ of the states versus $b$. (f) The real part of the instability growth rate, $\mathrm{Re}(\protect\delta )$, versus $b$. Green triangles correspond to (a) and (b), and red circles to (c) and (d). Parameters are $\protect\theta =0.1$, $V_{0}=-5$ and $G=-20$.}
\label{fig:three}
\end{figure}

On top of the simple few-beam self-trapped states, like the conservative 2D nonlinear systems~\cite{HS}, our $\mathcal{PT}$-symmetric systems can also support complex multi-beam ones built of up to 12 beams, see Fig.~\ref{fig:three}. Due to the interaction between individual beams, their intensity is larger at the center of the structure, the intensity difference gradually vanishing with the increase of propagation constant $b$. Near the interface in the truncated system, there are no beams located in the linear medium, while the near-interface beams become stronger, see Fig.~\ref{fig:three}(c,d). Power $P$ increases with propagation constant $b$ for both the in-bulk and surface self-trapped states, see Fig.~\ref{fig:three}(e). Results of the linear-stability analysis for these states, displayed in Fig.~\ref{fig:three}(f), reveal a common stability region for the in-bulk and surface modes, at intermediate values of propagation constants $b$.

\begin{figure}[htb]
\centerline{\includegraphics[width=1.0\columnwidth]{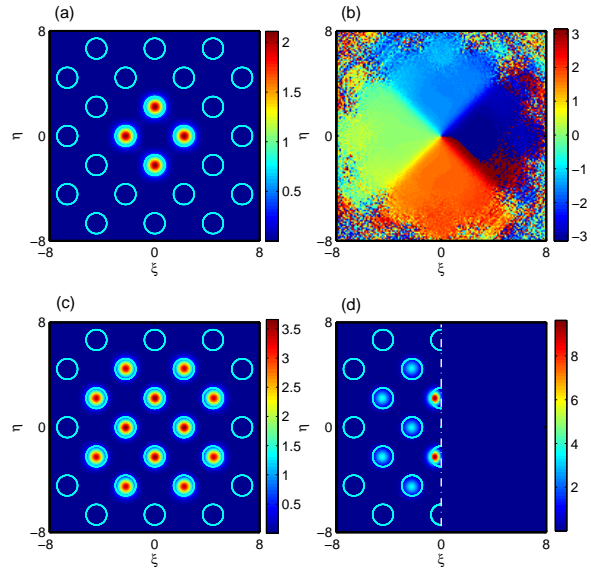}}
\caption{(color online) Long-distance propagation with $5\%$ noise. (a) The density profile at $\protect\zeta=500$ evolves from the the in-bulk vortex self-trapped nonlinear waves in Fig.~\protect\ref{fig:two} (a). (b) The corresponding phase distribution for (a). (c) The density profile at $\protect\zeta=500$ evolves from the in-bulk multi-beam self-trapped mode in Fig.~\protect\ref{fig:three} (b). (d) The density profile at $\protect\zeta=500$ evolves from the surface multi-beam self-trapped mode in Fig.~\protect
\ref{fig:three} (d).}
\label{fig:four}
\end{figure}

By simulating the beam propagation with $5\%$ random noise, we have verified the stability of the vortex modes, as shown in Fig.~\ref{fig:two}, and of multi-beam ones, see Fig.~\ref{fig:three}. In particular, Fig. \ref{fig:four}(b) demonstrates that the phase distribution of the input vortex mode keeps the phase-winding structure. Thus, the direct simulations corroborate the predictions of the linear-stability analysis.

In conclusion, we have found several novel species of in-bulk and surface self-trapped states in 2D Kerr-nonlinear optical systems with $\mathcal{PT}$-symmetric PLs (photonic lattices). These include stable in-bulk localized vortices and surface half-vortices (\textquotedblleft horseshoes"). The self-trapped modes are related to truncated nonlinear Bloch waves, the surface modes being linked with the truncated in-bulk ones. Along with the basic vortex and half-vortex states, which are built, respectively, of four and three constituent beams. The stable multi-beam self-trapped states, composed of up to 12 constituents, have been found too. The formation of these surface modes results from the interplay of the surface effects, nonlinearity, and the $\mathcal{PT}$-symmetry. Due to the surface-enhanced reflection, the discrete diffraction is stronger in the direction perpendicular to the interface than in the direction parallel to it~\cite{Kartashov,Rosberg,Suntsov2006,Szameit,Wang}, therefore the surface modes feature stronger nonlinearity, which is necessary to balance the diffraction.

\section*{Acknowledgments}
This work was supported by the National Basic Research Program of China under grants Nos. 2012CB821305, 2010CB923200 and 2013CB922403, the National Natural Science Foundation of China under grants Nos. 11374375, 11204043, 11274399 and 61078027, and the Ph.D. Programs Foundation of Ministry of Education of China under grant Nos. 20120171110022. B.A.M. appreciates hospitality of the Sun Yat-Sen University (Guangzhou, China).


\end{document}